\documentclass{article}
\usepackage[utf8]{inputenc}
\usepackage{amsmath}
\usepackage[charter]{mathdesign} 
\usepackage{hyperref}
\usepackage{algorithm2e}
\usepackage{graphicx}
\graphicspath{ {./figs/} }

\usepackage{listings}
\usepackage{tabularx}
\usepackage{booktabs}
\usepackage[labelfont=bf,format=plain,justification=raggedright,singlelinecheck=false]{caption}

\usepackage[table]{xcolor}

\def\ker{\widehat{f}}
\def\bw{w_{1:\infty}}
\def\uqtheta{\dot{\theta}}
\def\extheta{\theta}
\def\gen{p_{0}}

\title{Uncertainty Quantification and the Marginal MDP Model}
\author{Blake Moya \& Stephen G. Walker \\ \\
Department of Statistics \& Data Science \\
University of Texas at Austin}
\date{March 2022}

\begin{document}

\maketitle

\begin{abstract}
The paper presents a new perspective on the mixture of Dirichlet process model which allows the recovery of full and correct uncertainty quantification associated with the full model, even after having integrated out the random distribution function.  
The implication is that we can run a simple Markov chain Monte Carlo algorithm and subsequently return the original uncertainty which was removed from the integration. This also has the benefit of avoiding more complicated algorithms which do not perform the integration step. Numerous illustrations are presented.

\end{abstract}

\vspace{0.1in}

\section{Introduction}

The most well known and widely used Bayesian nonparametric model is the mixture of Dirichlet process (MDP) model.
The model yields a density $f(y)$ which is an infinite mixture of some \textit{kernel} density $\ker(y \mid \theta)$, and is of the form
\begin{equation} \label{mixture}
f(y) = \sum_{j = 1}^{\infty} w_{j} \, \ker(y \mid \uqtheta_{j}),\text{ where }\sum_{j = 1}^{\infty} w_{j} = 1.
\end{equation}
Here the $w_{1:\infty}$ are the weights and the $\uqtheta_{1:\infty}$ are a distinct set of $\theta$ parameters coming from some diffuse distribution. 
Due to its flexibility, the kernel is typically chosen to be normal, written as $N(y \mid \theta)$ with each $\theta_{j} = (\mu_{j}, \sigma^{2}_{j})$, representing the mean and variance of the mixture component.
It is prudent here to explain an alternative representation of the mixing proportions and weights as a single sequence. 
We may have some exchangeable sequence $(\extheta_j) = (\extheta_{1}, \extheta_{2}, \dots)$ which includes tied values.
The distinct values in this sequence will be written as $(\uqtheta_{j})$ and as such form an independent and identically distributed sequence.
The ties in $(\extheta_{j})$ construct the weights in the sense that
\begin{equation} \label{ndef}
N^{-1}(n_{N}^{1}, n_{N}^{2}, \ldots) \overset{a.s.}{\to} (w_{1}, w_{2}, \ldots) \quad \mbox{as} \quad N \to \infty
\end{equation}
where $n_{N}^{j}$ is the number of $(\extheta_{1}, \ldots, \extheta_{N})$ equal to $\uqtheta_{j}$.
In a Bayesian framework, a prior distribution $\gen(w, \uqtheta)$ is assigned for $\bw$ and $\uqtheta_{1:\infty}$ (or equivalently $\gen(\extheta)$ for $\extheta_{1:\infty}$) which, given the choice of kernel $\ker(y \mid \theta)$, is tantamount to imposing a prior directly onto $f(y)$.

The most commonly chosen and original such prior is the Dirichlet process, $DP(G \mid \alpha, G_{0})$ (see \cite{Ferguson1973}) where $\alpha$ is a scale parameter and $G_0$ is a distribution which serves as the prior expectation of the random $G = \sum_{j = 1}^\infty w_{j} \, \delta_{\uqtheta_{j}}$.
We can also write
$$
G = \lim_{N \to \infty} N^{-1}\sum_{j = 1}^{N} \delta_{\extheta_{j}}.
$$
In this way, and which becomes useful to us later, is that $G$ is essentially generating an infinite sequence of parameters $\extheta_{1:\infty}$ and, with the possibility of ties, making it possible to impute the mixing proportions $w$ for each \textit{unique} mixing parameter $\uqtheta$ in the sequence.
A detailed description of this distribution is provided in section (\ref{modelbackground}).


The infinite dimensional nature of the model renders it difficult to estimate.
The model was introduced in \cite{Lo1984}, though at that time the technical abilities to solve the prior to posterior problem were not available; see for example \cite{Kuo1986}. 
It was only with the advances in sampling based approaches to estimating posterior distributions that it became a standard model. 
The original Gibbs sampler was developed by \cite{Escobar1988} and later by \cite{Escobar1994}. 
Since then, numerous papers have contributed to the estimation of the model via the proposal of various MCMC algorithms. 
See \cite{Hjort2010} for a fairly recent review.  
The original Gibbs sampler was clever in that it removed the random distribution $G$ and hence turns the model into a finite one so that the samplers can be made exact. 
On the other hand, the removal of the random distribution takes away a large amount of the uncertainty associated with the full model. 
That is, of the $\extheta_{1:\infty}$ generated from the Dirichlet process, it is only a subset of the full sequence $\extheta_{1:n}$ that are estimated (where $n$ is the size of the data $y_{1:n}$).
Note that due to exchangeability the starting point of the estimated $n$-length sequence is irrelevant and is simply taken to be $1$.

While the predictive density $p(y_{n + 1} \mid y_{1:n})$ is available, the $\extheta_{n+1:\infty}$ that were integrated out have taken with them a substantial amount of uncertainty. 
Thus, questions of interest about the distribution of the population $y_{1:\infty}$ are unable to be answered.
More sophisticated algorithms retain the infinite mixture parameterized by $\uqtheta_{1:\infty}$ in the model. 
A finite approximation was introduced in \cite{Ishwaran2002} while \cite{Walker2007}, \cite{Papas2008} and \cite{Kalli2011} all develop algorithms which, while they retain $\uqtheta_{1:\infty}$ within the model, only need to sample a finite number of parameters within each iteration of a Markov chain Monte Carlo sampler to have the correction target posterior distribution. 
The downside is that the algorithms are quite complicated to implement, compared with that of \cite{Escobar1994}.

A further problem of the reliance on the parameters for inference, namely the mixing proportions and the component parameters, is the so--called label switching and identifiablity problems; see \cite{Stephens2000} for the former and \cite{Ferguson1983} for the latter. 
The first problem is an identifiability issue which is well known and can be difficult to solve; often requiring post MCMC processing.
The second problem often renders posterior summaries using the parameters as misleading, since there will be many parameter regions which all yield effectively an identical likelihood. See also \cite{Mena2015}.

A slightly different version of the mixture model 
is presented in \cite{Richard1997} and given by
\begin{equation} \label{finmixture}
f(y \mid k) = \sum_{j = 1}^{k} w_{j,k} \, \ker(y \mid \uqtheta_{j})
\end{equation}
with prior distributions being allocated to the number of  components $k$ and, conditional on $k$, the $(w_{1:k,k}, \uqtheta_{1:k})$. 
This model potentially looks simpler than (\ref{mixture}) though its estimation is complicated by the need to introduce a reversible Metropolis sampler for the number of components, due to a dimension change as $k$ increases or decreases. 
The theory for this is to be found in \cite{Green1995} and it is to be noted that the sampler described in \cite{Richard1997} is quite complicated.
Further, it also produces output similar to the Dirichlet process; see \cite{Green2001}.

In this paper we show how to undertake exact posterior inference 
with the marginal MDP model.
The algorithm is quite
simple and requires only a minor addition to the \cite{Escobar1994} algorithm. 
In particular, this addition is very fast and can utilize parallel computing. 
We can do this from the output of the simplest of the MDP algorithms (that of \cite{Escobar1988} and \cite{Escobar1994}) and obtain arbitrarily accurate results without the need for any new and/or complicated reversible jump algorithm. 
The necessary insight comes from the recent paper of \cite{Fong2022} who promote a fundamental alternative approach to the thinking of posterior distributions. 

The original idea on which \cite{Fong2022} is based is to be found in \cite{doob49}. 
Following $n$ samples in a Bayesian framework one has a predictive $p(\xi_{n + 1} \mid \xi_{1:n})$. 
Here $\xi$ can either represent data samples or hidden samples.
Sampling future observations and updating the predictive iteratively to collect an arbitrarily large sample $\xi_{1:M}$ yields a random distribution function $F(\xi)$, being the limit of the empirical distribution as the sample size grows. 
This random distribution can be taken as a distribution drawn from the posterior. 

Specifically, for the Dirichlet process, this is equivalent to showing that the limit of empirical distribution functions from a general P\'{o}lya--urn scheme is with probability one a random distribution chosen from a Dirichlet process prior (i.e. $\extheta_{1:\infty} \sim DP(\alpha, G_0)$ can be iteratively generated via an easily specified P\'olya--urn update rule). 
This was demonstrated by \cite{Black1973}, though it could have been deduced from Doob's paper \cite{doob49}.

This result does not necessarily need to start from the first observation. 
If the P\'olya--urn scheme is started after having an observed sample of size $n$ then the limit of the sequence is now a realization from the Dirichlet process posterior, conditioned on the observed sample. 
Consequently, from the \cite{Escobar1994} sampler, which at each iteration provides the first $n$ variables of an infinite P\'olya--urn sequence ($\extheta_{1:n}$), we can generate a realization from the posterior by continuing the sequence.
It is as straightforward as this.
And with a realization from the posterior, all forms of inference become possible. 
Further, there are no identifiability problems as the main inferential tool is targeting the distribution rather than the parameters.



\begin{table}
\caption{Notation used in this article}
\begin{tabularx}{\textwidth}{@{}p{0.2\textwidth}X@{}}
\toprule
  $f(y)$ & The nonparametric distribution of $y$ \\
  $\ker(y \mid \theta)$ & The kernel for a mixture distribution \\
  $\Theta$ & The parameter space for $\theta$ \\
  $\extheta_{1:M}$ & An exchangeable sequence of $M$ kernel parameters in $\Theta^{M}$ \\
  $\uqtheta_{1:M}$ & A set of $M$ unique i.i.d. kernel parameters in $\Theta^{M}$ \\
  $w,\;w_{1:M}$ & A set of $M$ mixing proportions in $\Delta^{M}$ (the $M$-simplex) \\
  $w_{j, k}$ & Entry $j$ of a mixing proportion set in $\Delta^{k}$ \\
  $G(\theta)$, $G$ & A distribution of the population $\extheta_{1:\infty}$ defined by $(w_{1:\infty}, \uqtheta_{1:\infty})$ \\
  $G_{M}(\theta)$, $G_{M}$ & A distribution of the finite population $\extheta_{1:M}$ defined by $(w_{1:k}, \uqtheta_{1:k})$ \\
  $\delta_{x}(z)$, $\delta_{x}$ & A Dirac point mass function at $x$ \\
\bottomrule
\end{tabularx}
\end{table}

\section{Model Background} \label{modelbackground}

In this section we go through the MDP model with full details and introduce our perspective on it by thinking about the random $G$ as a random infinite exchangeable sequence $\extheta_{1:\infty}$. 
The marginal MDP model removes the section $\extheta_{n+1:\infty}$ when $G$ is integrated out and hence to recover the full uncertainty 
we must recover these parameters.

\subsection{The MDP Model}

We start off by looking at mixture models in general. 
As previously described, a mixture model $f(y)$ with number of components $k$ with kernel density $\ker(y \mid \theta)$ is formulated as in \ref{finmixture}.
The prior distribution on $f(y)$, once the kernel and degree have been fixed, generally has the form
\begin{equation*}
    p_{0}(f(y)) = \gen(w_{1:k}, \uqtheta_{1:k}) = \text{Dir}(w \mid p_{1}, \dots, p_{k}) \prod_{j = 1}^{k} G_{0}(\uqtheta_{j} \mid \eta)
\end{equation*}
where (with a slight abuse of notation since we write $G_{0}$ as a distribution function rather than a density) $G_{0}$ is a distribution on the kernel parameter space $\Theta$, parameterized by some hyper--parameter $\eta$.

We can understand this model via a truncated MDP model.
For some $N$ let $\extheta_{1:N}$ be a truncated infinite sequence. Then we can define
$$
f(y) = N^{-1} \, \sum_{j = 1}^{N} \ker(y \mid \theta_{j})
$$
which can also be written as
$$
f(y) = \sum_{j=1}^{k} w_{j, k} \, \ker(y \mid \uqtheta_{j}),
$$
where $w_{j, k} = N^{-1}n_{N}^{j}$ (as in (\ref{ndef})) and $k$ is the number of distinct $\uqtheta_{1:k}$. 
This is a MDP motivated version of the \cite{Richard1997} model; essentially a truncated MDP model, similar in spirit to \cite{Ishwaran2002}.
This representation underlies a convenient style of Gibbs samplers for mixture models in which one samples $(w_{1:k}, \uqtheta_{1:k}, \extheta_{1:n})$ iteratively from their full conditionals.
Notably, it is possible during data generation to have some $\uqtheta_{j}$ not appear in $\extheta$. 
That is, some of the $k$ components may not end up being responsible for the generation of any of the $n$ data points. 
Such components are generally called ``empty".

The general mixture model is written as
$$
f(y)=\int \ker(y \mid \theta) \, dG(\theta)
$$
where $G$ is assigned a prior distribution. 
The MDP model uses the likelihood form of (\ref{mixture}) and a Dirichlet process prior for $G$.
Here we focus on the notion that $G$ is also represented by the infinite sequence $\extheta_{1:\infty}$. 
We write this as $DP(\extheta_{1:\infty} \mid \alpha, G_{0})$. 
This can be represented as a limiting form of the joint density for a P\'{o}lya--urn scheme generated sequence, 
\begin{equation} \label{polyaurn}
\text{P\'{o}l}(\extheta_{1:M} \mid \alpha, G_{0}) = \prod_{i = 1}^{M} \frac{\alpha \, G_{0}(\extheta_{i}) + \sum_{j < i} \delta_{\extheta_{j}}(\extheta_{i})}{\alpha + i - 1},
\end{equation}
where $\delta_\theta$ is the Dirac point mass at $\theta$.
As $M \to \infty$ the sequence $\extheta_{1:M}$ converges with probability one to a sample from the Dirichlet process (i.e. $\extheta_{1:\infty} \sim DP(\alpha, G_{0})$, see \cite{Black1973}),
\begin{equation*}
    \lim_{M \to \infty} \text{P\'{o}l}(\alpha, G_{0}, M) = DP(\alpha, G_{0}).
\end{equation*}
Estimating the full model relies on the latent variable driven likelihood function given by
\begin{equation} \label{fullmodel}
l(w, \uqtheta) = \prod_{i = 1}^{n} w_{d_{i}} \, \widehat{f}(y_{i} \mid \uqtheta_{d_{i}}),
\end{equation}
where $d = \{d_{1}, \ldots, d_{n}\}$ is a vector of latent allocation variables. 
Here the $(w, \uqtheta)$ are taken according to a Dirichlet process model; so the $\uqtheta_{1:\infty}$ are i.i.d. $G_0$ and the $w_{1:\infty}$ are constructed via $w_{1} = v_{1}$ and, for $j > 1$, $w_{j} = v_{j} \prod_{i < j}(1 - v_{i})$ with each $v_{j}$ a $\mbox{Beta}(1, \alpha)$ distributed random variable.

Sampling this version of the model using (\ref{fullmodel}) is difficult due to the lack of a normalizing constant for the mass function of each $d_{i}$ which can take any value in $\{1, \ldots, \infty\}$. 
All the complexities of the algorithms associated with the full model are about handling this problem. 
Approximating by truncating the support for each of the $d_{i}$ at some large number (see \cite{Ishwaran2002}) is one possible attempt.
Other ideas include introducing further latent variables which naturally restrict the choice for each $d_{i}$ to a finite set (see \cite{Walker2007}, \cite{Papas2008} and \cite{Kalli2011}).

Even after one has samples from the posterior one has the so--called label switching problem to deal with. 
This can be understood in simple terms; for example, that the values of $\uqtheta_{1}$ and $\uqtheta_{2}$ sampled in one iteration of the Markov chain may be swapped in a different iteration. 
The labels representing a possible set of the $y_{i}$ can change across iterations of MCMC algorithm. 
This is not the only identifiability problem associated with the full model (\ref{fullmodel}). 
As \cite{Ferguson1983} has pointed out, the parameters $w$ and $\uqtheta$ are not identifiable as there are many different settings for the parameters which recover the likelihood. 
For example, though by no means the most serious case, a standard normal density can be recovered with two standard normal density functions with arbitrary choices of weights.

We now look at the marginal MDP model and our perspective on it.

\subsection{The Marginal MDP Model}

To motivate the marginal model it is written in hierarchical form;
$$
f(y_{i} \mid \theta_{i}) = \ker(y_{i} \mid \theta_{i}), \quad i = 1, \ldots, n,
$$
and $(\theta_{1:n} \mid G)$ are i.i.d. from $G$. 
The $G$ is then assigned the Dirichlet process prior. 
Integrating out the $G$ results in the marginal model whereby the first part of the hierarchical model remains the same but the second part is now that the $(\theta_{1:n})$ are distributed as (\ref{polyaurn}) with $M = n$.

We will now write the MDP in an alternative way by relying on the equivalence between the distribution $G$ and the population it characterizes $\extheta_{1:\infty}$.
We write the MDP model as 
$$
p(y \mid G) = p(y_{1:n} \mid \extheta_{1:\infty}) \, p_{0}(\extheta_{1:\infty}),
$$
where $p_0$ is the model for the infinite exchangeable P\'olya--urn
sequence and $p(y_{1:n}\mid \extheta_{1:\infty})$ only requires the first $n$ entries in the sequence (which we can claim due to exchangeability) and becomes $p(y_{1:n} \mid \extheta_{1:n})$. 
The $p_0(\extheta_{1:\infty})$ component can be written as
\begin{equation} \label{decompose}
p_0(\extheta_{1:n})\,\,p_0(\extheta_{n+1:\infty} \mid \extheta_{1:n}).
\end{equation}
This representation (\ref{decompose}) forms the key insight.
So integrating out $G$ to form the marginal model is nothing more than simply ignoring the $p_0(\extheta_{n+1:\infty} \mid \extheta_{1:n})$ part of the model.
This becomes very clear by writing 
\begin{equation} \label{key}
p(y \mid G) = p(y_{1:n} \mid \extheta_{1:n}) \, p_0(\extheta_{1:n}) \, p_0(\extheta_{n+1:\infty} \mid \extheta_{1:n}).
\end{equation}
Hence, the posterior for $\extheta_{1:n}$ can be written as
\begin{equation} \label{ourrep}
p(\extheta_{1:n} \mid y_{1:n}) \, p_{0}(\extheta_{n+1:\infty} \mid \extheta_{1:n})
\end{equation}
where 
$$
p(\extheta_{1:n} \mid y_{1:n}) \propto p_{0}(\extheta_{1:n}) \, \prod_{i = 1}^{n} \widehat{f}(y_{i} \mid \extheta_{i}).
$$
The full posterior is then
\begin{equation} \label{fullposterior}
p(\extheta_{1:\infty} \mid y_{1:n}) \propto p(\extheta_{1:n} \mid y_{1:n}) \,\, p_0(\extheta_{n+1:\infty} \mid \extheta_{1:n}).
\end{equation}
So we view the marginal model not as an integrated model, but rather the sidelining, or the ignoring of the $p_0(\extheta_{n+1:\infty}\mid\extheta_{1:n})$
part of the model.

The representations (\ref{ourrep}) and (\ref{fullposterior}) provide clearly the insight into our approach. 
We can implement a marginal MDP sampling algorithm to get the posterior samples $\extheta_{1:n}$ and then ``return" the (temporarily) removed $[\extheta_{n+1:\infty} \mid \extheta_{1:n}]$
by simply sampling the completion of a P\'olya--urn scheme.
While surely we are unable to sample the entire infinite sequence, we can select an arbitrarily large number $M$ at which to truncate the sample. 
Critically, our method allows for a specification of an easily interpretable truncation error for the samples.
We detail the algorithm that accomplishes this in section \ref{sec:margmdp}. 
Our contention is that the removed samples (i.e. $\extheta_{n+1:\infty}$) are very easy to put back to recover the posterior uncertainty implied by the original model, and as a consequence we only need to conduct a simple marginal MDP sampling algorithm. 

\subsection{Alternative Exchangeable Sequences}

Although we focus on the P\'olya--urn scheme, in practice, any exchangeable sequence $(\extheta_{1:\infty})$ can be used. 
In one direction, we could use the well known Pitman--Yor exchangeable sequence (\cite{Pitman1997}) which generalizes, with an extra single parameter, the P\'olya--urn sequence. 
Beyond this are the Gibbs type exchangeable sequence (see \cite{Favaro2013}). 
These are often expressed in terms of the partition distributions which highlight the feature of ties among a sequence of $n$ samples. 
For the Gibbs type exchangeable sequences, the exchangeable sequence can be understood from the ties distribution. That is, for any $1 \leq k \leq n$,
$$
P_{n,k}(n_{1}, \ldots, n_{k}) = V_{n,k} \prod_{j = 1}^{k} (1 - \sigma)_{(n_{j} - 1)},
$$
where $k$ is the number of distinct $\extheta_{1:n}$, the $n_{j}$ are the number of corresponding ties, $\sigma \in [0, 1)$, $\extheta_{(m)} = \extheta(\extheta + 1) \ldots (\extheta + m - 1)$ and the $V_{n,k}$ satisfy
$$
V_{n,k} = (n - k \sigma) \, V_{n+1, k} + V_{n+1, k+1}.
$$
While these are associated with a nonparametric model, on the other hand, it is possible to write down the joint density of an exchangeable sequence based on a parametric model. 
For example, if
$$
[\extheta_{i} \mid \mu] \overset{iid}{\sim} \, N(\nu, \sigma^{2}) \quad \mbox{and} \quad \mu \sim N(0, \tau^{2})
$$
then $p_0(\extheta_{1:n})$ is multivariate normal distribution with mean 0 and inverse covariance matrix
$\Sigma^{-1} = \Omega / \sigma^{2}$ where $\Omega$ is an $n \times n$ matrix with diagonal elements $1 - a$ and off diagonal elements $-2a$ where
$a = \tau^{2} / (\sigma^{2} + n \tau^{2})$. 
It is not difficult to also find $p(\theta_{1:n})$ when $\sigma$ is also unknown; since then it would be a multivariate Student--$t$ distribution. 
The joint distribution will typically be available for conjugate models.

\section{P\'{o}lya Completion for the Marginal MDP} \label{sec:margmdp}

In this section we present the details of the algorithm, which can be viewed as an extension to the Gibbs sampler of \cite{Escobar1994}. 
This sampler takes five inputs: a data vector $y_{1:n}$, a mixture kernel $\ker(y \mid \theta)$, a base measure $G_{0}(\theta \mid \eta)$, and fixed values or hyperprior specifications for base measure parameter set $\eta$ and concentration parameter for the Dirichlet process $\alpha$. 
For some total number of sampling iterations $T$, the sampler outputs
$$
(\extheta_{1:n}^{(t)}, \eta^{(t)}, \alpha^{(t)})_{t = 1:T}.
$$
The Gibbs sampler has the following iterative update steps:

\begin{align*}
    \extheta_{1:n} \sim{}& p(\extheta_{1:n} \mid y_{1:n}, \eta, \alpha) \\
    \eta \sim{}& p(\eta \mid \extheta_{1:n}) \\
    \alpha \sim{}& p(\alpha \mid \extheta_{1:n}).
\end{align*}
Our extension can be viewed a the addition of one further update:
\begin{align*}
    \extheta_{n+1:M} \sim{}& p(\extheta_{n+1:M} \mid \extheta_{1:n}, \eta, \alpha).
\end{align*}
Because the extended sequence is not included in the full conditional updates for the other parameters, we can actually reserve this sapling step until the collection of $(\extheta_{1:n}^{(t)}, \eta^{(t)}, \alpha^{(t)})_{t = 1:T}$ is already complete. 
With these we can sample $\extheta_{n+1:M}^{(t)}$ and then construct the random distribution functions $G_{M}^{(t)}$, which are the corresponding empirical distribution functions of the samples. 
Not only does this mean that the extension can be applied in parallel for each iteration $t$ (though we find it can be implemented so quickly that this is unnecessary), it also means that any previously collected samples can be retroactively completed without augmenting or rerunning previous software.

The sampling of the $\extheta_{n+1:M}^{(t)}$ is completed as follows; we take $\extheta_{n+1}^{(t)}$ from the density proportional to
$$
p(\cdot \mid \extheta_{1:n}^{(t)}, \ldots) \propto \alpha^{(t)} \, G_{0}\big(\cdot \mid  \eta^{(t)}\big) + \sum_{i = 1}^{n} \delta_{\extheta_{i}^{(t)}}(\cdot),
$$
and subsequently, and in general,  we take $\theta_{m+1}^{(l)}$ given $(\theta_{1:m}^{(l)})$ from the density proportional to

$$
p(\cdot \mid \extheta_{1:m}^{(t)}, \ldots) \propto \alpha^{(t)} \, G_{0}\big(\cdot \mid  \eta^{(t)}\big) + \sum_{i = 1}^{m} \delta_{\extheta_{i}^{(t)}}(\cdot).
$$
Note that the sequence $(\theta_{n+1:M}^{(t)})$ is itself a P\'{o}lya--urn sequence with starting parameters $\alpha^{(t)} + n$ and 
$$
G_{n}^{(t)}(\cdot) = \frac{\alpha^{(t)} \, G_{0}(\cdot \mid \eta^{(t)}) + \sum_{i = 1}^{n} \delta_{\extheta_{i}^{(t)}}(\cdot)}{\alpha^{(t)} + n}.
$$
The natural question to ask now is how many samples one should take in order to construct $G^{(t)}$, and during our answer we will drop the superscript $t$. 
Now $G_{M} = M^{-1} \sum_{j = 1}^{M} \delta_{\extheta_j}$; yet an alternative approach is to use the representation of \cite{Sethuraman1994}. 
This involves taking a set $\phi_{1:M}$ to be i.i.d. from $G_{n}$ and taking $v_{1:M}$ to be i.i.d. $\mbox{Beta}(1, \alpha + n)$. Then we can take, using some $M$ to be determined,
$$
G_{M}(\theta) = \sum_{j = 1}^{M} w_{j}\,\delta_{\phi_j}(\theta)
$$
where $w_{1} = v_{1}$ and, for $j > 1$, $w_{j} = v_{j} \prod_{i < j}(1 - v_{i})$. 
We would want to take $M$ large enough so that $W_{M} = \sum_{j = 1}^{M} w_{k} > 1 - \epsilon$ for some specified $\epsilon > 0$. 
It is easy to check that this happens when 
\begin{equation} \label{criterion}
\prod_{j = 1}^{M}(1 - v_{j}) < \epsilon.
\end{equation}
So one plan is to keep sampling until (\ref{criterion}) holds. 
The implication is that $M$ is randomly driven, and a nice result appearing in \cite{Tardella1998} indicates that $M$ is a $1 + \mbox{Pois}(-\alpha \log\epsilon)$ random variable. 
To see this note that (\ref{criterion}) is equivalent to 
$$
-\sum_{j = 1}^{M} \log(1 - v_{j}) \geq -\log(\epsilon)
$$
and the $-\log(1 - v_{j})$ are i.i.d. exponential random variables with means $1 / (\alpha + n)$. 
The Poisson number then follows from an elementary result in probability theory. 
In practice, we can take $M$ to to be the Poisson quantile up to some additional error $\upsilon$ (i.e. $M = 1 + \mbox{Pois}(\upsilon; -\alpha\log\epsilon)$) to guarantee a bound on running time.

The random distribution from the posterior is 
$$
G_{M}(\theta) = \sum_{j = 1}^{M} w_{j} \, \delta_{\phi_{j}}(\theta) + w_{M+1} \, \delta_{\phi}(\theta),
$$
where $w_{M+1} = 1 - W_{M} = 1 - \sum_{j = 1}^{M} w_{j}$ and $\phi$ is taken from $G_{n}$.
The corresponding random density function representing a sample density function from the posterior is 
$$
f^{(t)}(y) = \sum_{j = 1}^{M} w^{(t)}_j \, \widehat{f}(y \mid  \phi_{j}^{(t)}) + w_{M+1}^{(t)} \, \widehat{f}(y \mid \phi^{(t)}).
$$
It is important to understand this is not a predictive density; which is why we have reverted to using the superscript $t$ for each iteration. 
While it is  similar in appearance to what could possibly be obtained using more sophisticated algorithms, such as \cite{Papas2008} and \cite{Kalli2011}, the key point is that for us the Markov chain is the simplest available and the extension to get the $f^{(t)}$ is essentially instantaneous even without the parallel computation which is available.
For general MDP models with base measure $G_{0}(\theta \mid \eta)$, the P\'{o}lya completion algorithm is given in Algorithm (\ref{alg:polya}).

With multiple samples $(f^{(t)})_{t=1:T}$ it is possible to achieve posterior summaries which have previously been elusive using the typical output from MCMC algorithms for the MDP model. 
This is a novel feature to MCMC algorithms and is of great significance; in the sense that each $f^{(t)}$ represents a possible true density, while the variation over the $t$ represents the uncertainty under the full MDP model. 

\SetKwComment{Comment}{/* }{ */}
\RestyleAlgo{ruled}
\begin{algorithm}
\caption{P\'{o}lya Completion for MDP}\label{alg:polya}
\KwData{$\alpha$, $G_{0}$, $\eta$, $\theta_{1:n}$, $\epsilon$, $\upsilon$}
\KwResult{$w_{1:M}$, $\phi_{1:M}$}
 Set $M \gets 2 + $Pois$\big(1 - \upsilon ; - (\alpha + n) \log(\epsilon)\big)$\;
 Draw $v_{1:M} \overset{iid}{\sim}$ Beta$(1, \alpha + n)$\;
 \For{$m$ in $\{1,\dots,M\}$}{
  \uIf{$m = 1$}{
   Set $w_{m} \gets v_{m}$\;
  }
  \uElseIf{$m \neq M$}{
   Set $w_{m} \gets v_{m} \prod_{j = 1}^{m} \left(1 - v_{j}\right)$\;
  }
  \Else{
   Set $w_{m} \gets 1 - \sum_{j=1}^{M-1}\phi_{j}(w)$\;
  }
  Draw $u \sim$ Unif$(0, \alpha + n)$\;
  \eIf{$u \leq \alpha$}{
   Draw $\phi_{m} \sim G_{0}(\phi \mid \eta)$\;
  }{
   Draw $\phi_{m}$ uniformly at random from $\theta_{1:n}$\;
   \uIf{$\phi_{m} = \phi_{k}$ for some $k < m$}{
   Set $w_{k} \gets w_{k} + w_{m}$\;
   Set $w_{m} \gets 0$\;
   }
  }
 }
 Remove $w_{k}, \phi_{k}$ for $\{k : w_{k} = 0\}$\;
\end{algorithm}

\section{Illustrations}

\subsection{Available Software}

We illustrate the methodology with the use of custom software written in R \cite{R2022} with core code written in C++ via Rcpp \cite{Eddelbuettel2011}. 
The open source code is available at
\url{https://github.com/blakemoya/MDPolya}, 
and this contains the functions \lstinline{mdp(data, k)} for conducting the marginal MDP Gibbs sampler of \cite{Escobar1994} given a data vector and a desired number of samples, \lstinline{polya(res_mdp, eps, ups)} for implementing Algorithm (\ref{alg:polya}) given output from the \lstinline{mdp} function, a desired level of accuracy, as well as functions for evaluating, plotting, and counting the modes of the resulting mixture densities. 
The \lstinline{polya} function has an additional method for \lstinline{dirichletprocess} object generated by the \lstinline{dirichletprocess} R package \cite{Ross2022} for access to the range of base measures implemented in that package. 

The functions all have useful default settings, with the default hyperparameters, burn-in, and thinning factor for \lstinline{mdp} set according to those used for the \lstinline{galaxies} demonstration in \cite{Escobar1994} (including additional arguments to fix $\alpha$, $\mu$, and/or $\tau$). 

Before proceeding to the specific illustrations, we briefly recall the model specification of \cite{Escobar1988} and \cite{Escobar1994}. 
The marginal Normal MDP model of \cite{Escobar1988} and \cite{Escobar1994} is given at the first level by
$$
[y_{i} \mid \extheta_{i}] \sim N(\cdot \mid \extheta_{i})
$$
for $i \in \{1, \dots, n\}$ and the $\extheta$ are generated according to a Dirichlet process with base measure $G_{0}$ taken to be a normal--inverse-gamma distribution $\mbox{NIG}(\cdot \mid \mu, \tau, s, S)$. 
Consequently, the first $n$ realizations $\extheta_{1:n}$ follow a P\'{o}lya--urn scheme, giving the second level of the model as
$$
\extheta_{1:\infty} \sim DP(G_{0}, \alpha) \;\; \to \;\; \extheta_{1:n} \sim \mbox{P\'{o}l}(G_{0}, \alpha, n) : G_{0} = \mbox{NIG}(\mu, \tau, s, S),
$$
and therefore the additional hyperpriors may be assigned to impose the following third level to the model:
$$
    \mu \sim{}  N(a,A) \quad
    \tau \sim{} \mbox{Inv-Ga}(w, W) \quad \mbox{and} \quad
    \alpha \sim{} \mbox{Ga}(c, C).
$$
In order to implement a Gibbs sampler, it is convenient that all the conditional distributions are of known form, and hence easy to sample. 
Indeed, all the necessary distributions to sample are provided in section 3 in \cite{Escobar1994}. 
Following an implementation of a Gibbs sampler for $T$ iterations, with burn--in and thinning iterations if desired, the output of interest is 
$$
(\extheta_{1:n}^{(t)}, \mu^{(t)}, \tau^{(t)}, \alpha^{(t)})_{t = 1:T}.
$$

The layout of the illustrations is as follows: In section \ref{subsec:gal} we will reproduce the \lstinline{galaxies} demonstration from \cite{Escobar1994} and then apply the P\'{o}lya completion algorithm to extend the results. 
We will compare the sampled densities as well as the resulting distributions over indicators of the cardinality of the underlying mixtures. 
In section \ref{subsec:synth} we will generate synthetic data from the prior used in \ref{subsec:gal} to compare the performance of the marginal and full MDP models when ground truth is known.
Finally, in section \ref{subsec:bench} we benchmark the software and reveal just how quickly the results of the aforementioned sections were obtained.


\subsection{The \lstinline{galaxies} data} \label{subsec:gal}

\subsubsection{Replication}

We illustrate the efficiency and speed of the algorithm by reproducing and extending the results of the \cite{Escobar1994} marginal MDP sampler on the popular \lstinline{galaxies} data set. 
The \lstinline{galaxies} data is a sample of $n = 82$ recorded velocities of distant galaxies in km/sec (scaled to Mkm/sec for stability) and was obtained from the \lstinline{MASS} R package \cite{Venables2002}. 
This is the data taken to be $y_{1:n}$ for the model given in Section (\ref{sec:margmdp}). We set the hyperparameters $\{c, C, a, A, w, W, s, S\} = \{1, 2, 20.8, 20.8, \frac{1}{2}, 50, 2, 1\}$, all in line with the original illustration except for $A = 20.8 \approx \text{Var}(y)$ which was originally undefined to induce an improper prior. 
The justification for the selection of the other hyperparameters can be found in \cite{Escobar1994}.
\footnote{Note that some of the Gamma distribution parameters in \cite{Escobar1994} are presented in halves (i.e. $s/2, S/2, w/2, W/2$). We also present them halved.} 
We ran the marginal MDP sampler for 2000 burn-in iterations and thinned the remaining samples by a factor of 150. With these values all set as defaults in the software, we were able to generate the samples with a single line.

\begin{lstlisting}[language=R]
res_mdp <- mdp(MASS::galaxies / 1000, 100)
\end{lstlisting}

\begin{figure}
    \centering
    \includegraphics[width=6cm,height=4cm]{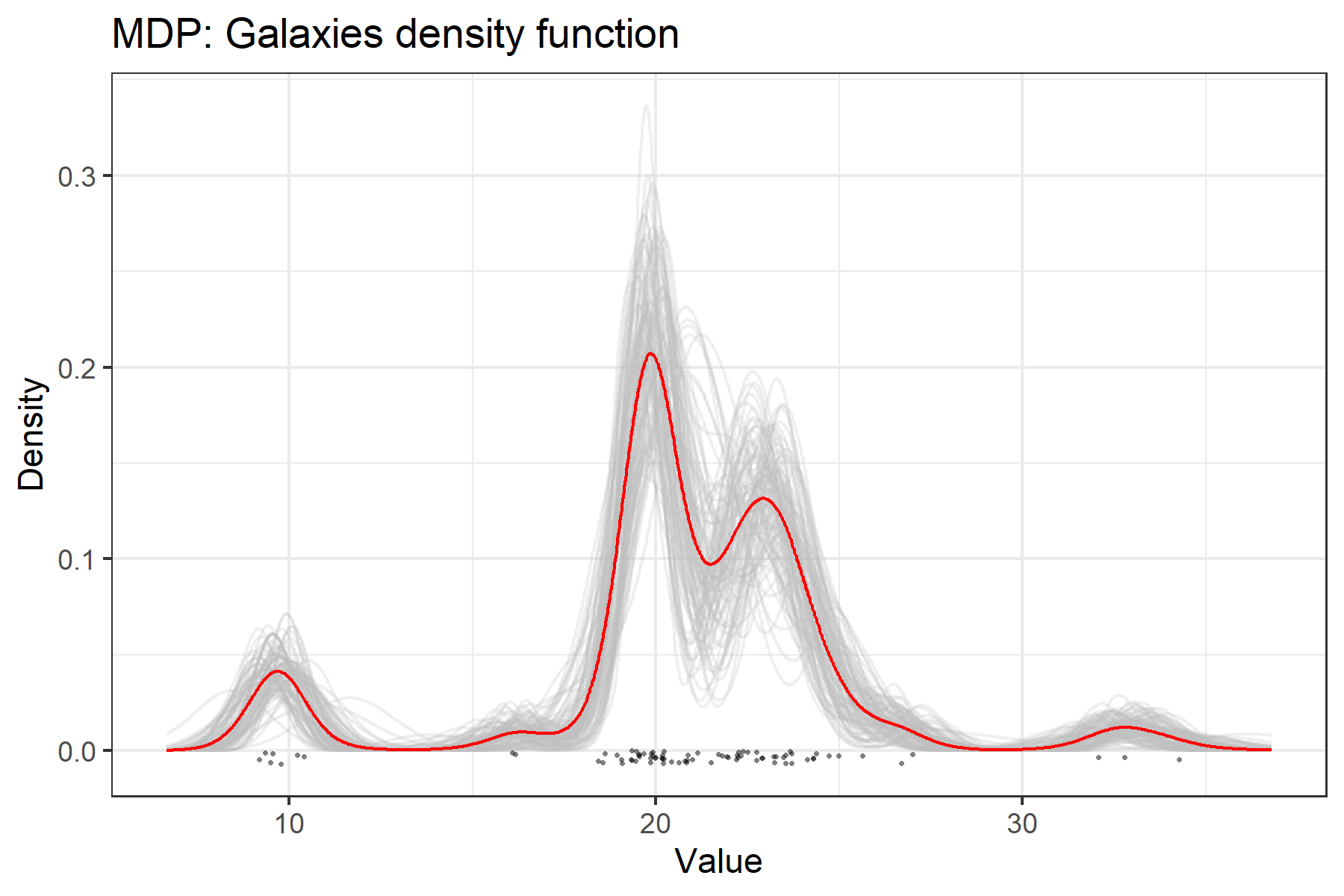}
    \includegraphics[width=6cm,height=4cm]{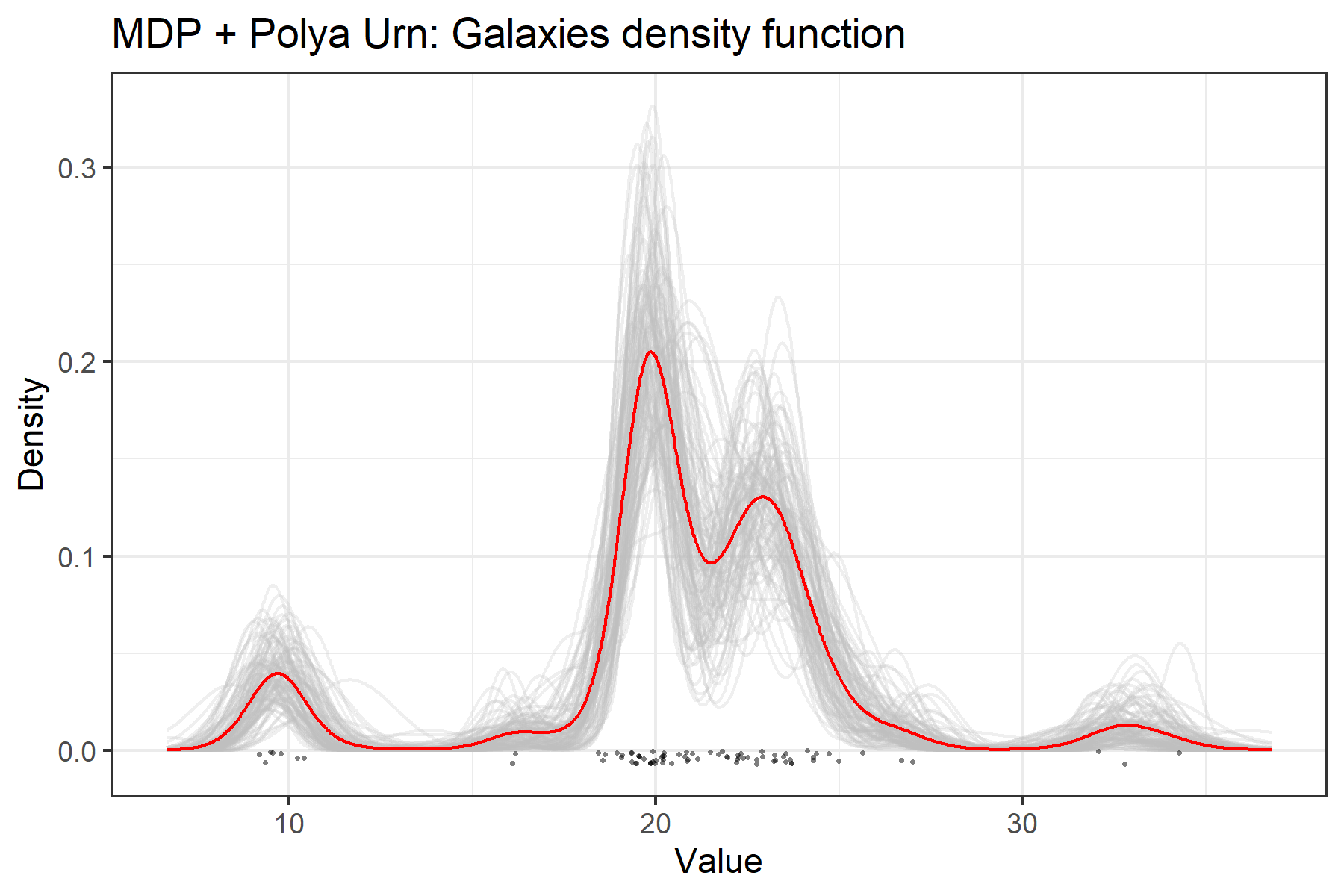}
    \caption{A comparison of the sampled MDP density functions before and after P\'{o}lya completion. Each individual sample is represented by a grey density, with the point--wise mean in red. The original $y_{1:n}$ are shown as black points.}
    \label{fig:galaxiespdf}
\end{figure}

\begin{figure}
    \centering
    \includegraphics[width=6cm,height=3cm]{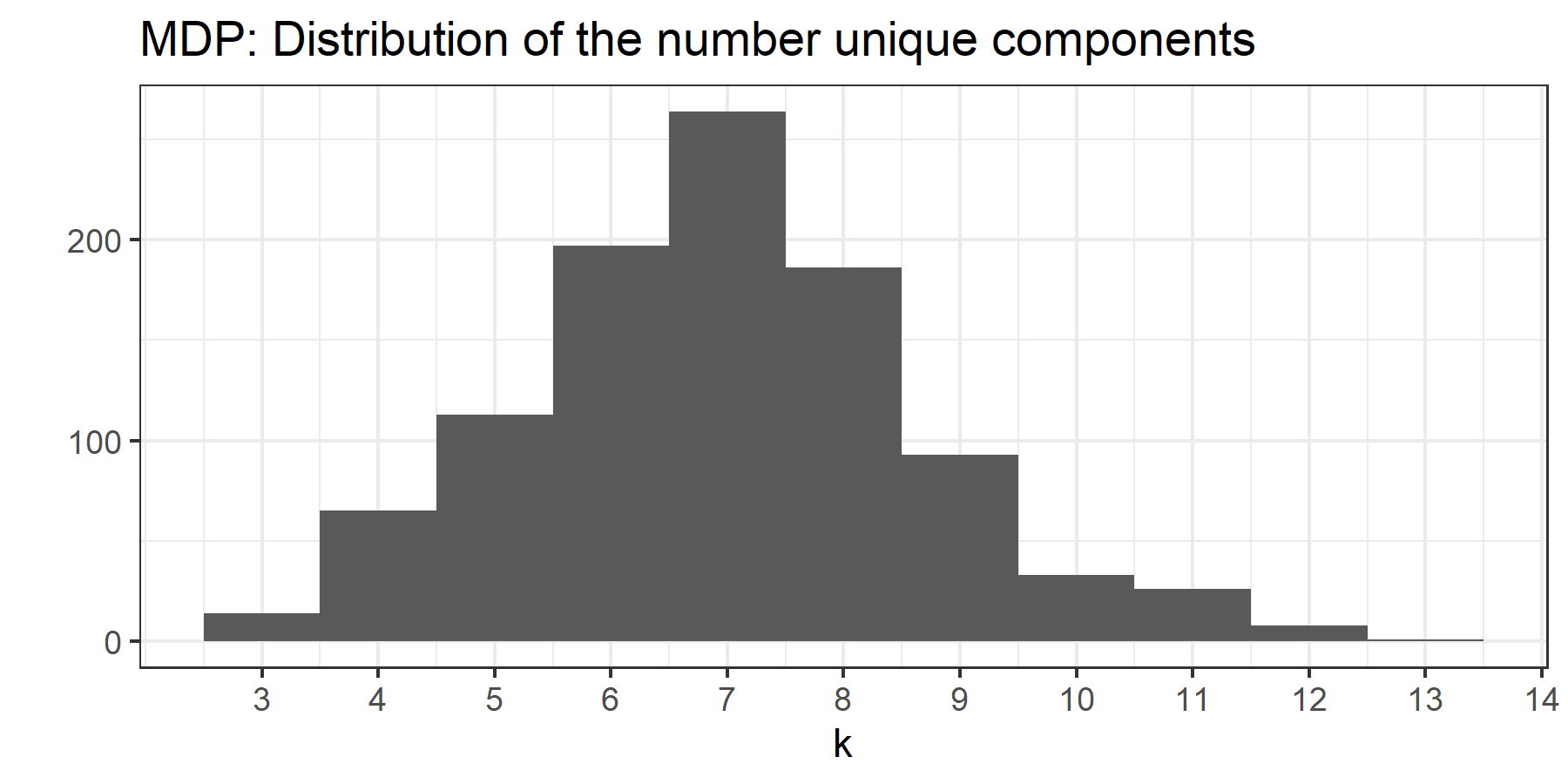}
    \includegraphics[width=6cm,height=3cm]{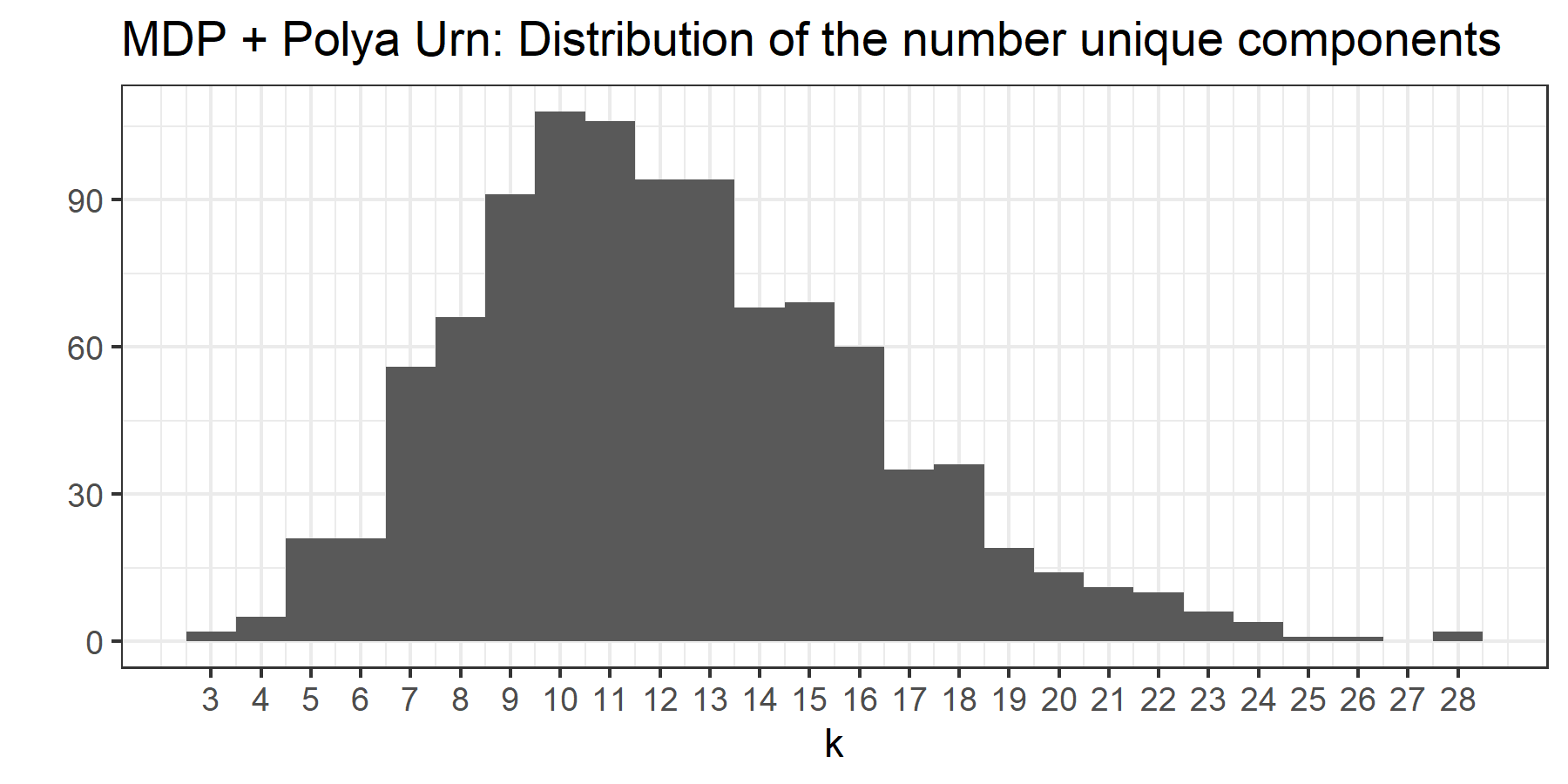}
    \caption{A comparison distribution over the number of mixture components of the sampled mixture densities before and after P\'{o}lya completion.}
    \label{fig:comps}
\end{figure}

The recorded samples $\extheta_{1:n}^{(t)}$ form the sampled density estimates via $p(y_{i} \mid \extheta_{1:n})$, and 100 such samples can be seen in the left-hand side of Figure \ref{fig:galaxiespdf}. 
As expected, the results resemble the original illustration of the algorithm quite well. 
Part of the initial illustration of the marginal MDP sampler was its use as a mixture deconvolution tool when the MDP is ``considered as a proxy for a finite mixture model", \cite{Escobar1994}, with an unknown number, $\kappa$, of components. 
Inference on $\kappa$ was done by assuming that each mixture component is a normal distribution and simply tallying the $k \in \{1, \dots, n\}$ unique values in $\extheta_{1:n}$ as output by the sampler. 
Because each mixture density is stored as a matrix of mixture parameters $\phi_{1:k}$, the number of components $k$ can be obtained with
\begin{lstlisting}
nk_mdp <- sapply(res_mdp$phi, nrow)
\end{lstlisting}
and the resulting distribution for 1000 marginal MDP samples is shown in Figure \ref{fig:comps}.

\subsubsection{Extension}

We may now extend the samples from the marginal MDP model into samples from the full model. 
Each marginal MDP sample can be fed independently (and in parallel) into Algorithm (\ref{alg:polya}) to yield a vector $\phi_{1:M^{(t)}}^{(t)}$ of component parameters and a vector of associated weights $w_{1:M^{(t)}}^{(t)}$. 
We use a default $\upsilon = 0.01$ such that on average $1 - \upsilon = 0.99$ of the $w_{1:M^{(t)}}^{(t)}$'s meet better than the desired error $\epsilon = 0.01$ 
(that is, about 99\% of the sampled mixtures are truncated only after the weights sum to $0.99$ or greater). 
We can obtain these results using the following single line of code from the software, and the densities produced by it are displayed on the right side of Figure \ref{fig:galaxiespdf}.

\begin{lstlisting}[language=R]
res_pol <- polya(res_mdp, 0.01, 0.01)
\end{lstlisting}

\begin{figure}
    \centering
    \includegraphics[width=6cm, height=4cm]{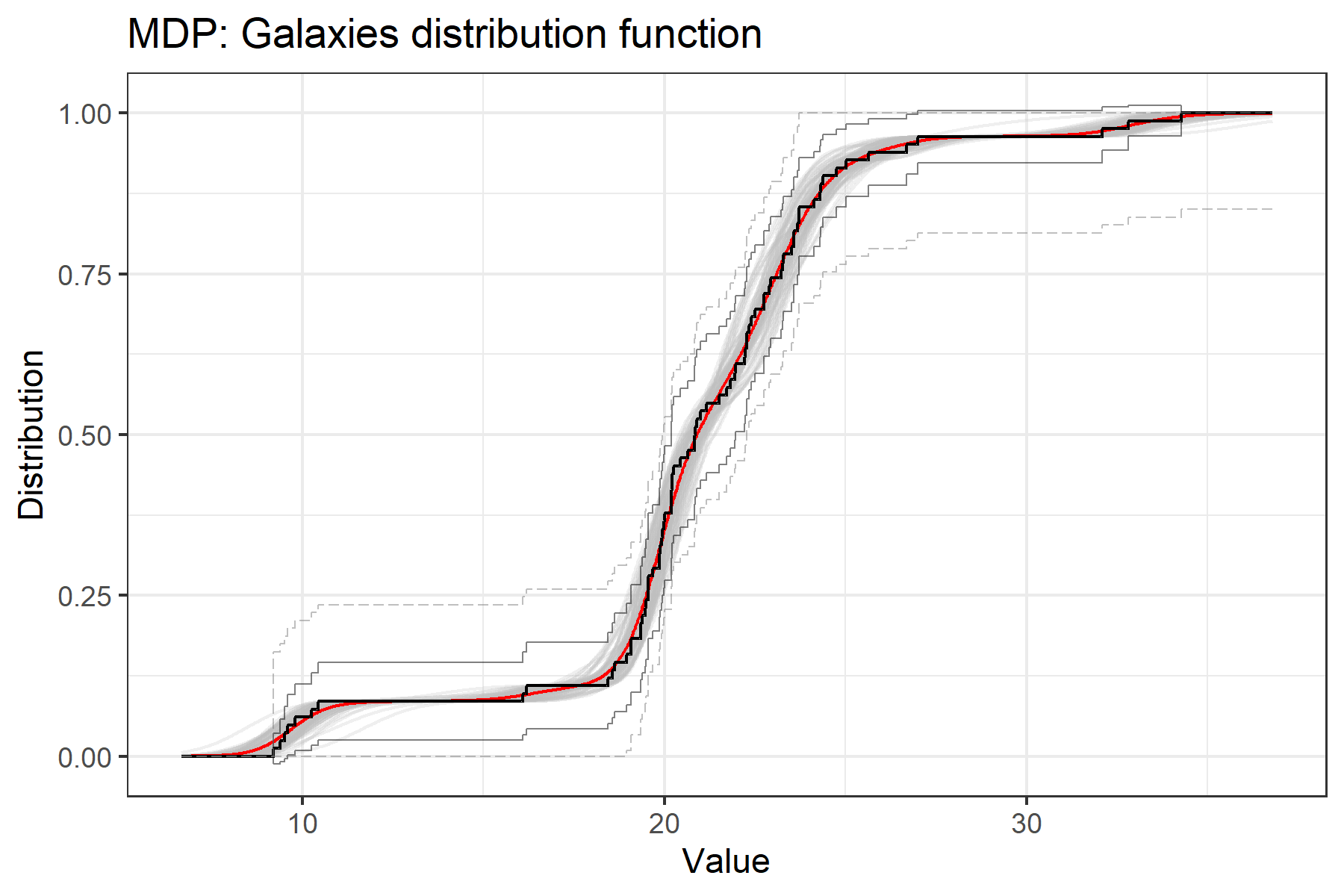}
    \includegraphics[width=6cm,height=4cm]{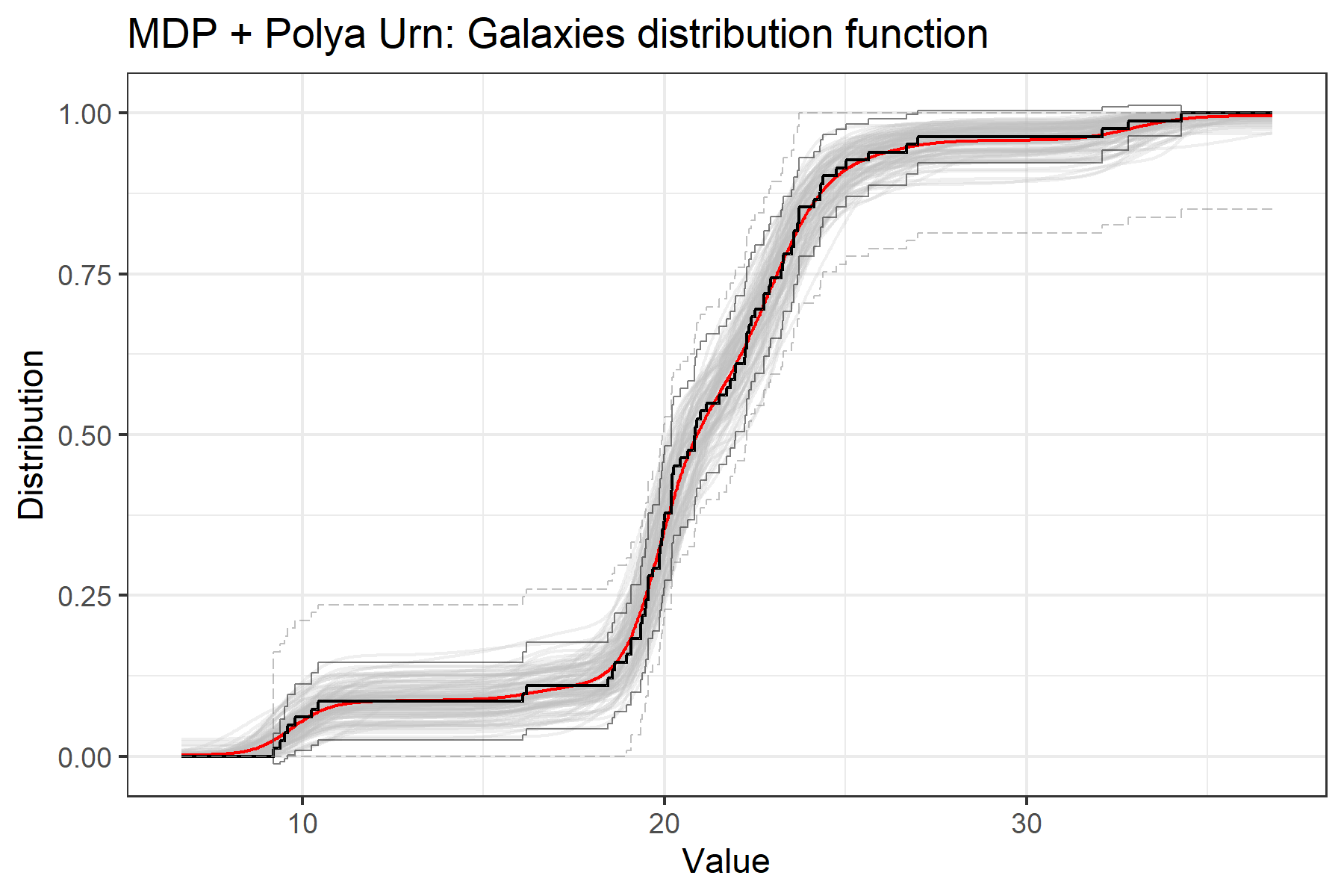}
    \caption{A comparison of the sampled MDP distribution functions before and after P\'{o}lya completion. Each individual sample is represented by a grey distribution, with the point--wise mean in red. The empirical distribution of $y_{1:n}$ is shown in black. The point--wise and simultaneous 95\% confidence bands are shown in solid and dashed grey, respectively.}
    \label{fig:galaxiescdf}
\end{figure}

Hence, we have captured the full uncertainty quantification associated with the full MDP model. 
The densities represented by each $\phi^{(t)}$ and $w^{(t)}$ look similar to the \cite{Escobar1994} counterpart densities, but with a notable amount of additional variability. 
The distribution functions reveal the difference much more clearly. 
In Figure \ref{fig:galaxiescdf} the distribution functions of the same samples are displayed with both point--wise and simultaneous 95\% confidence bands. 
It is clear here how marginalization of the MDP model discards the uncertainty contained in mixture components that are not represented by any one of the finite data points. 
The primary utility of the MDP model is its ``nonparametric" characterisation, so it seems apt that the full MDP model fits almost perfectly along the two sets of nonparametric confidence bands shown with it. 
The marginal model density function from \cite{Escobar1994} show dramatically reduced variation; a clear symptom of the removal of the random distribution from the model.

The number of mixture components $k$ explicitly calculated in Algorithm (\ref{alg:polya}) now no longer has the same meaning as that for the marginal model since we no longer consider the nonparametric infinite mixture as a proxy for a parametric finite one. 
The utility of the an infinite mixture of normals, again, is precisely in the construction of a nonparametric distribution. 
Representing a nonparametric mixture density as a finite mixture of normals needlessly restricts our interpretation of what a ``mixture component" can be in the first place. 
For instance, using the MDP model to fit a unimodal skewed distribution will yield a $\theta_{1:n}$ which contains $k > 1$ normal mixture components, many of which contributing mass to the heavier tail of the distribution. 
In cases such as this, it makes less sense to characterize the mixture by it's additive normal components than by its modes 
(nevertheless, we show the distribution over $[k \mid \epsilon = \upsilon = 0.01]$ in Figure \ref{fig:comps} for comparison).

\begin{figure}
    \centering
    \includegraphics[width=6cm,height=3cm]{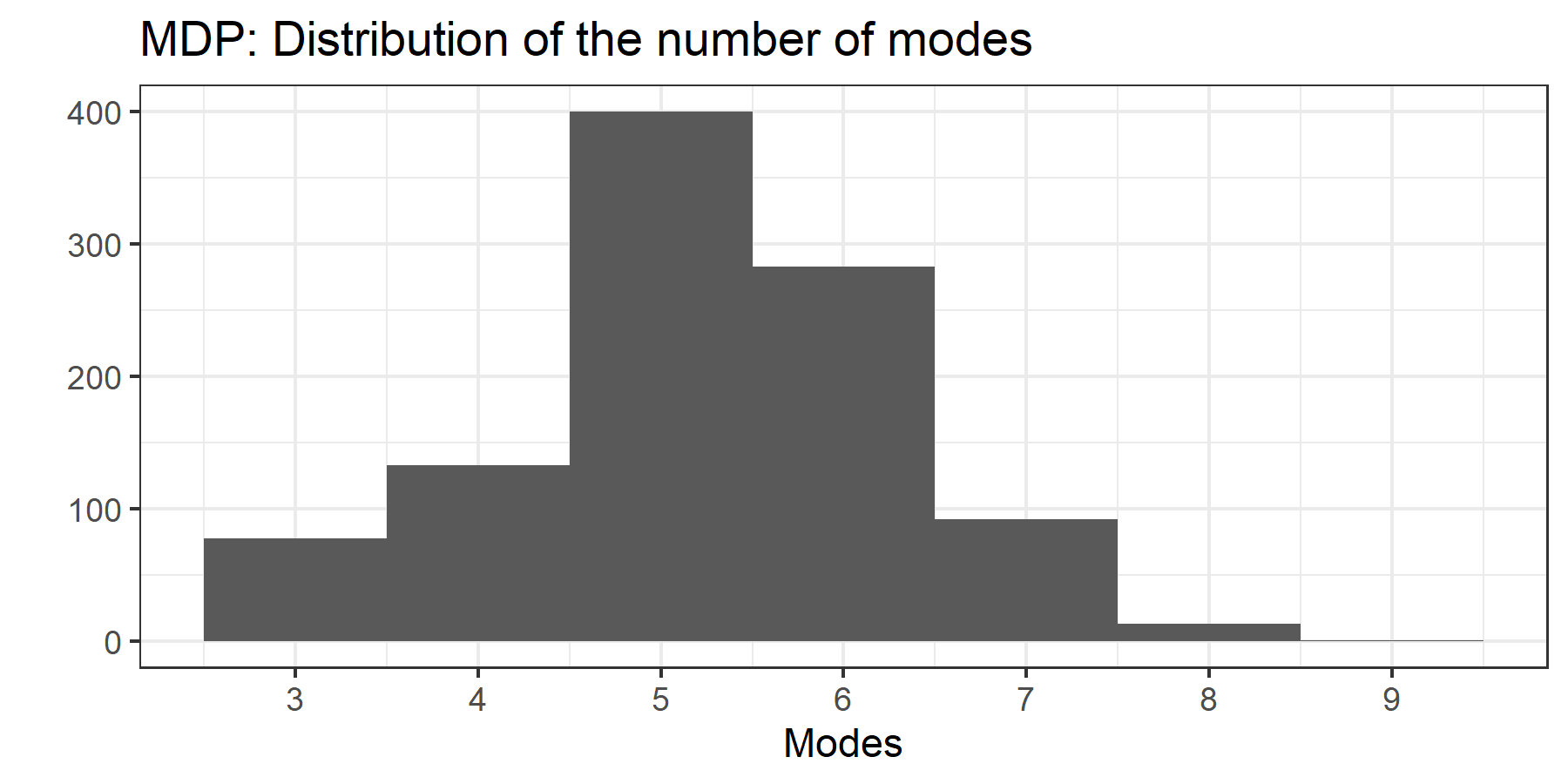}
    \includegraphics[width=6cm,height=3cm]{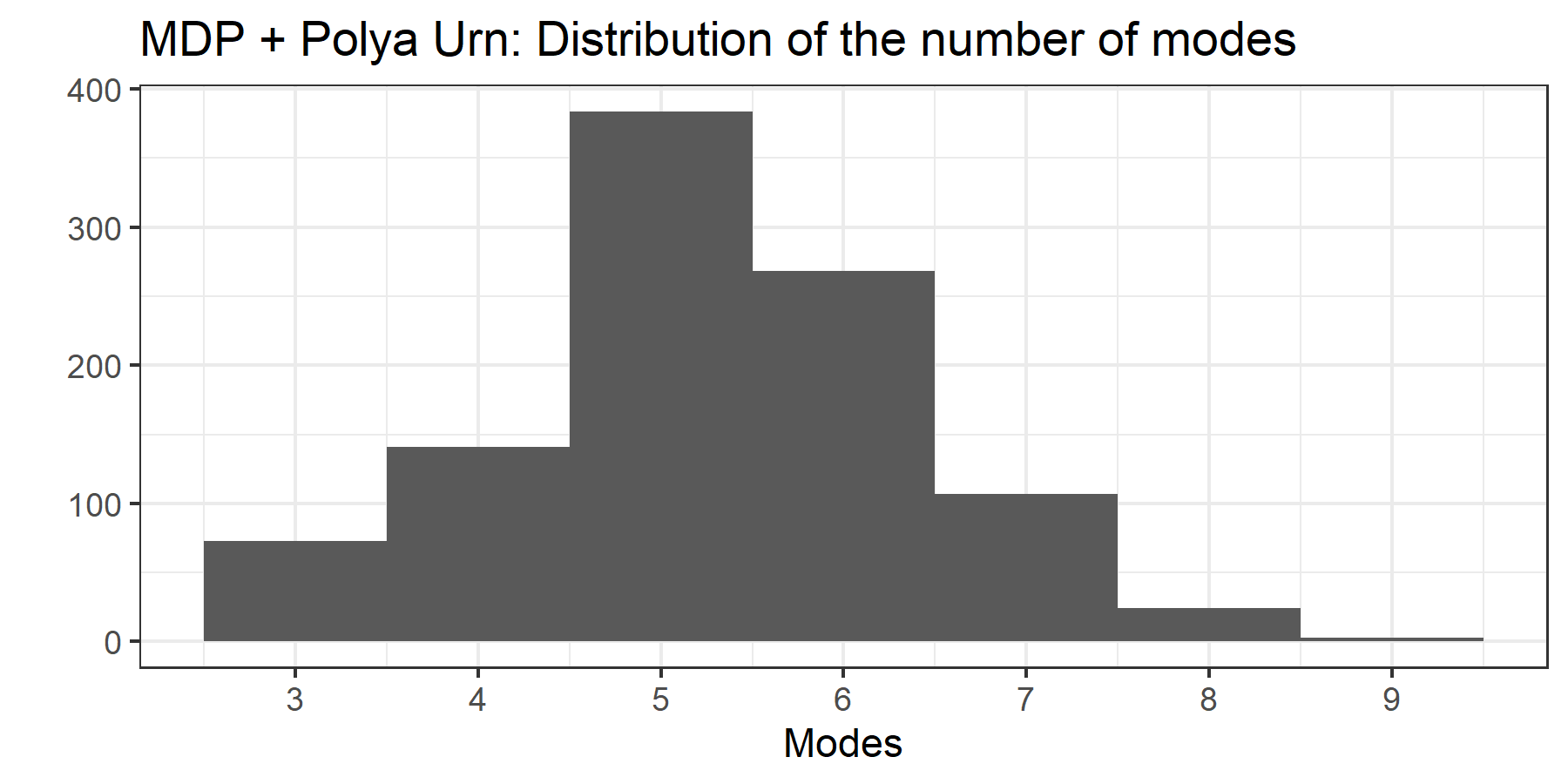}
    \caption{A comparison distribution over the number of modes of the sampled mixture densities before and after P\'{o}lya completion}
    \label{fig:modes}
\end{figure}

We used a grid based method like the one referenced in Escobar \& West to count the number of modes within the range of the data $\hat{m}$ of the sampled densities from the marginal MDP model and the full MDP model via P\'{o}lya completion (for a more precise calculation mode--counting algorithm , see \cite{Carreira2000}). 
The code to obtain the location of those modes for each sampled density from the \lstinline{mdp} or \lstinline{polya} functions is simply:

\begin{lstlisting}
m_mdp <- modes(res_mdp)
m_pol <- modes(res_pol)
\end{lstlisting}

and vectors of the mode counts can be obtained with:

\begin{lstlisting}
nm_mdp <- sapply(m_mdp, length)
nm_pol <- sapply(m_pol, length)
\end{lstlisting}

Figure \ref{fig:modes} compares the posterior distribution of $\hat{m}$ from 1000 samples of the marginal MDP and full models. 
The number of components vastly overestimates the number of modes, as many low-weight components can be introduced to better characterize the density, even if they don't change gross characteristics such as the number of modes. 
The distribution over the modes in the full MDP model agrees quite well with the marginal MDP, with only slightly more variation.

With access to the full model, we can now quantify uncertainty over more interesting statistics. 
By using  $\phi_{1:M}$ to evaluate the nonparametric density over a fine grid, we can use trapezoidal integration 
to approximate any functionals of the sampled distributions to a high degree of accuracy. 
In the following section, we will do just that with simulated data so that we can view our results with the true values known.

\subsection{Simulation Study} \label{subsec:synth}

We  use the \cite{Sethuraman1994} method to generate the $\phi$ and $w$ from the prior distribution and generate synthetic data with knowledge of the true generating distribution as well as which mixture component each data point was generated from. 
With the same prior hyperparameters as used for the \lstinline{galaxies} demonstration, we can generate MDP distributions and sample 
$
k_{i} \sim \mbox{Multi}(w_{1}, \dots, w_{M})
$ 
and $[y_{i} \mid k_{i}] \sim N(\cdot \mid \phi_{k_{i}})$. 
Since we know which of the $k$ components each $y_{i}$ was conditionally drawn from, we can reconstruct a truncated version of the MDP distribution from only the components which actually generated a $y_{i}$ and with weights proportional to the number of $y_{i}'s$ each generated. 
This represents the marginal MDP on which the inference is concentrated in the marginal MDP model.

\begin{figure}
    \centering
    \includegraphics[width=6cm,height=4cm]{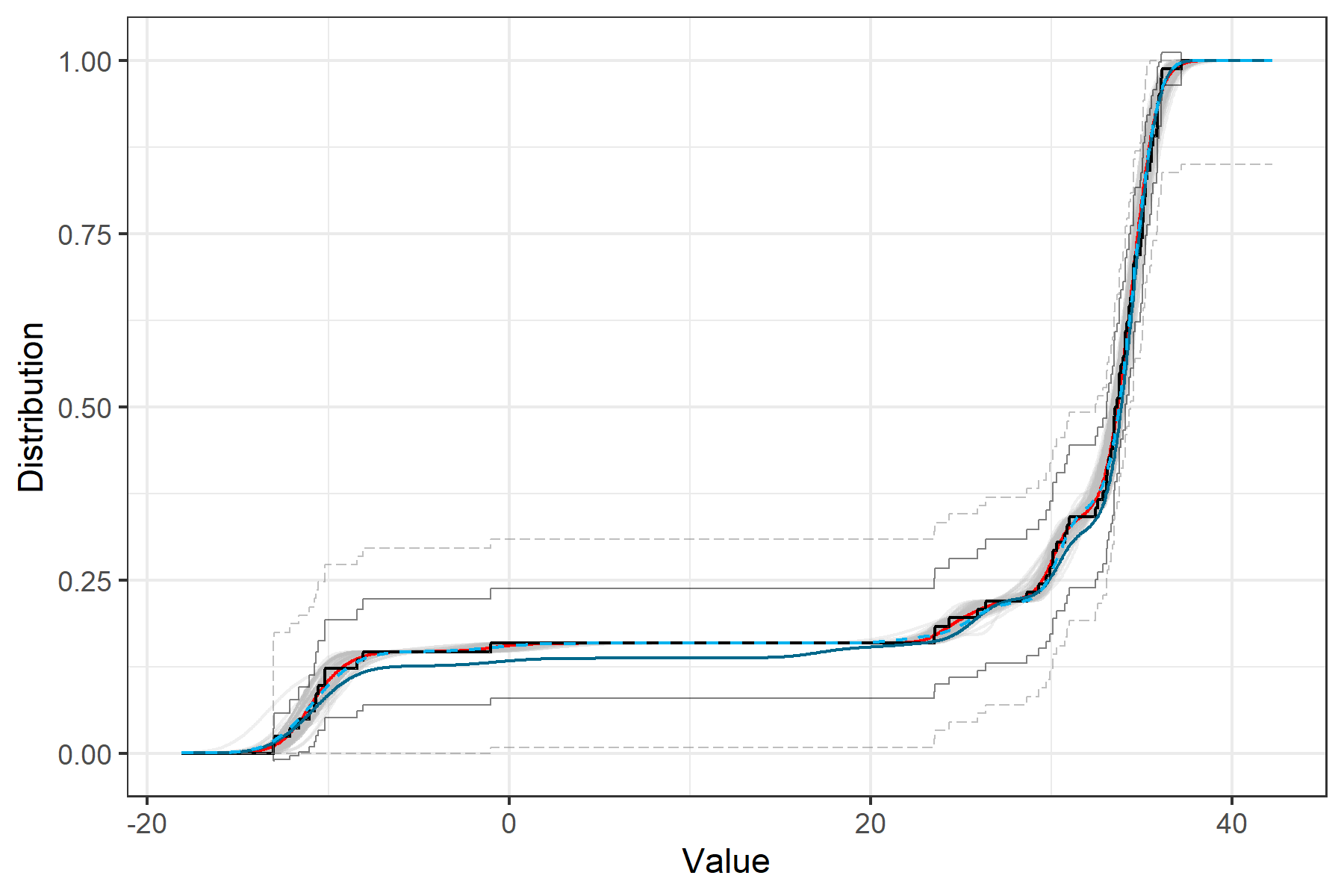}
    \includegraphics[width=6cm,height=4cm]{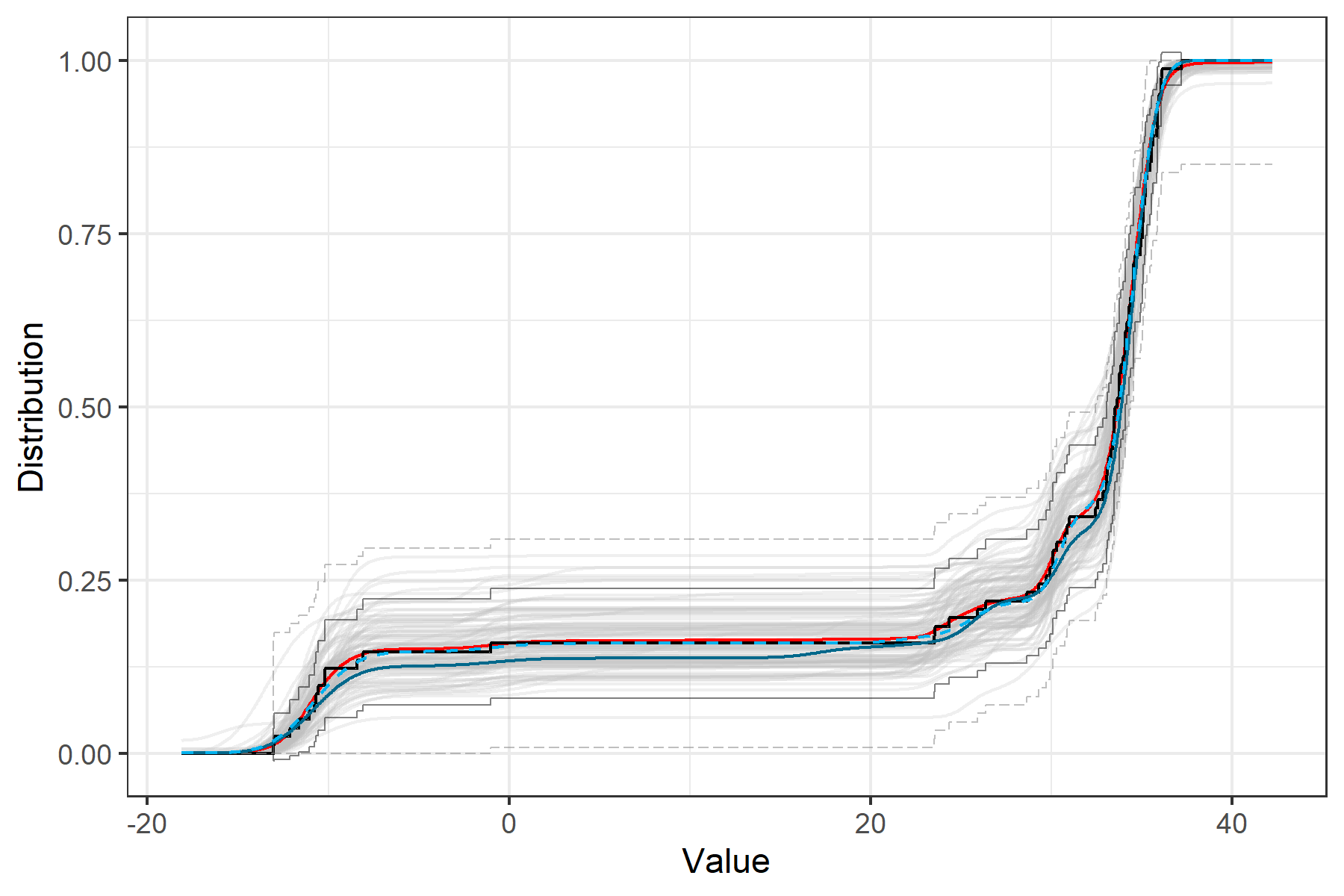}
    \caption{A comparison similar to Figure \ref{fig:galaxiescdf}, this time with synthetic data. In addition to the information represented in Figure \ref{fig:galaxiescdf}, there is also a solid dark--blue line representing the true full MDP distribution that generated the data and a dashed light--blue showing the truncated marginal distribution implied by $\extheta_{1:n}$.}
    \label{fig:sim1}
\end{figure}

Figure \ref{fig:sim1} shows the result of such a sampling, where samples of size $n=82$ were drawn from distributions sampled from the prior induced by the hyperparameters set as in Section \ref{subsec:gal}. 
The marginal MDP samples stick tightly to the empirical distribution of the sample, and often fail to capture the true or truncated distributions that generated the sample, especially in multimodal distributions. 
After P\'{o}lya completion, however, this weakness is resolved.

\begin{figure}
    \centering
    \includegraphics[width=6cm,height=6cm]{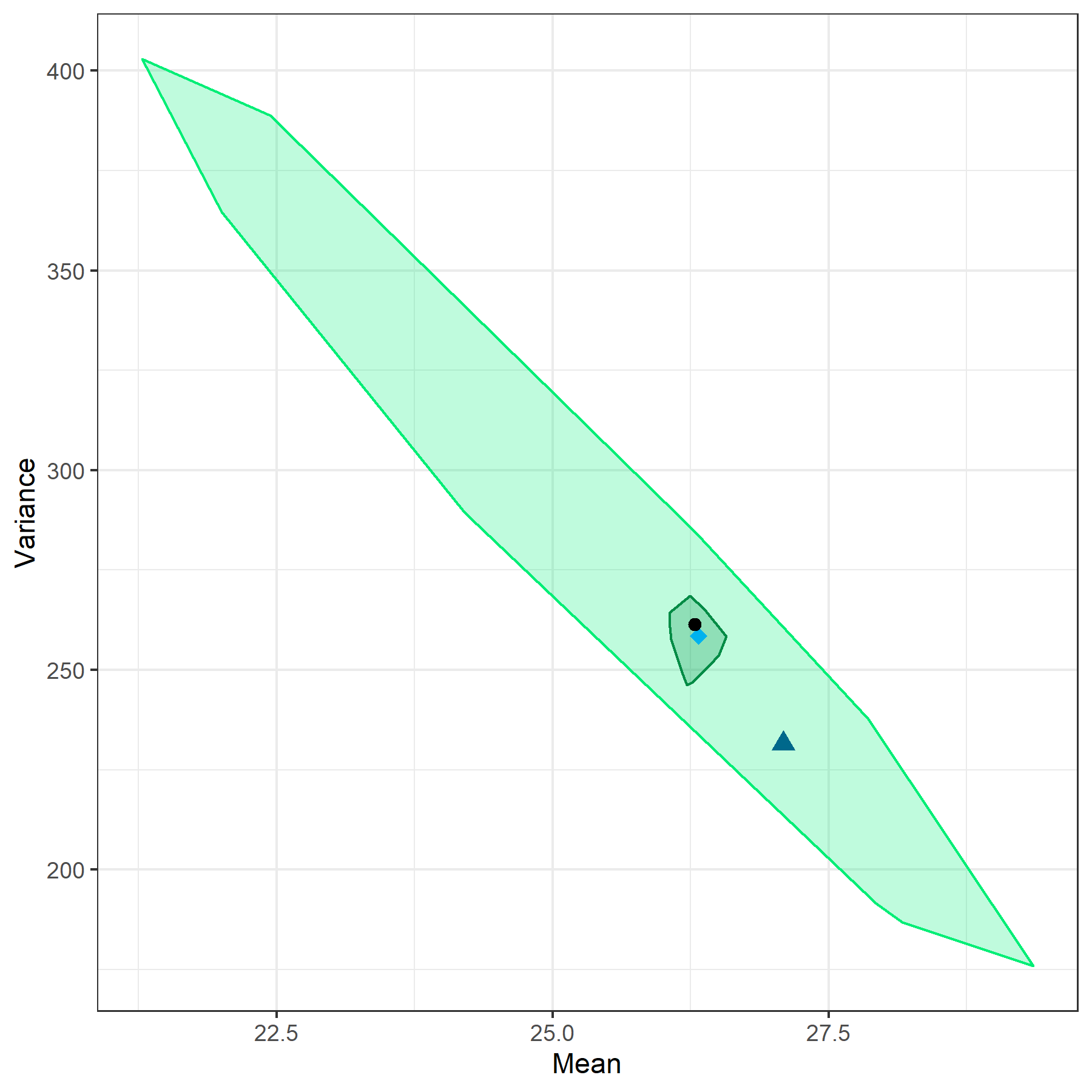}
    \caption{The posterior distribution over the first two central moments over the distribution in Figure \ref{fig:sim1}. A region containing 95\% of 1000 sampled moments from the marginal MDP model is indicated in dark--green. A similar region for the same samples after P\'{o}lya completion is shown in light--green. The empirical mean and variance is represented the black circle, the true population moments by the dark--blue triangle, and the moments of the truncated marginal distribution by the light--blue diamond.}
    \label{fig:moms}
\end{figure}

Further, we use a trapezoidal integration technique to calculate the first two (central) moments for each distribution and this obtain a distribution over the population mean and variance for both the marginal and full MDP models. 
In Figure \ref{fig:moms} we display these distributions with the true population mean and variance, as well as the empirical.
We find that the full model, as expected, captures the full uncertainty in the moment estimates and thus frequently contains the true values in it's 95\% confidence region.
More interestingly, we find that the additional uncertainty imbued by the P\'{o}lya completion algorithm seems to be scaled based on the distance of the empirical mean and variance to the prior expectation of the mean and variance.

\subsection{Algorithmic Efficiency} \label{subsec:bench}

\begin{figure}
	\centering
	\includegraphics[width=12cm,height=5cm]{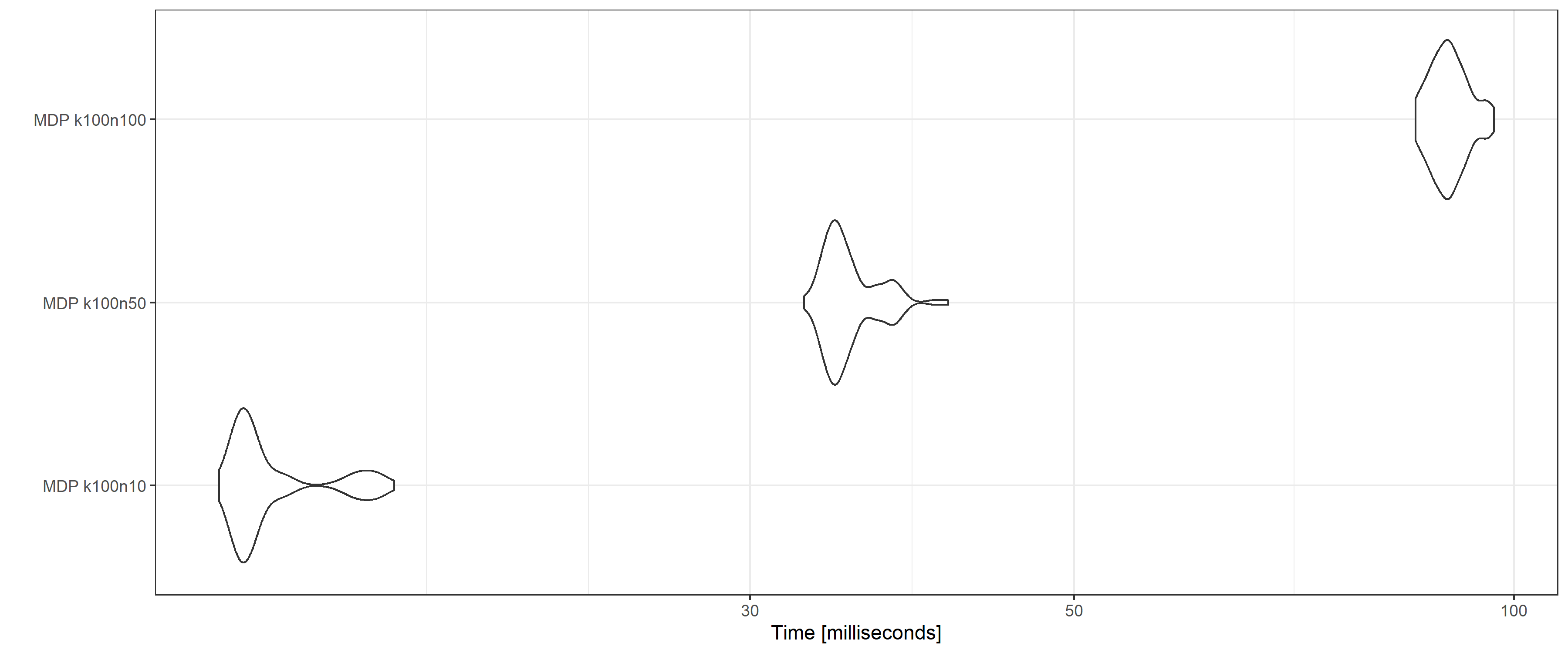}
	\caption{A violin plot of the running time for 100 runs of \lstinline{mdp(n, k = 100)} for three different values of $n$.}
	\label{fig:bench_mdp}
\end{figure}

The speed of the algorithm is remarkable, especially due to the simplicity compared to the complicated MCMC schemes which hitherto have been necessary to accomplish the same thing. 
The Gibbs sampler itself is very fast due to its simple construction, taking less than 1 millisecond per iteration on datasets with a sample size $n = 100$. 
Distributions of the running time of the \lstinline{mdp} function over various values of sample size $n$ and iterations $k$ are shown in Figure \ref{fig:bench_mdp}. 
Note that these benchmarks do not use burn--in or thinning iterations. 

\begin{figure}
	\centering
	\includegraphics[width=12cm,height=5cm]{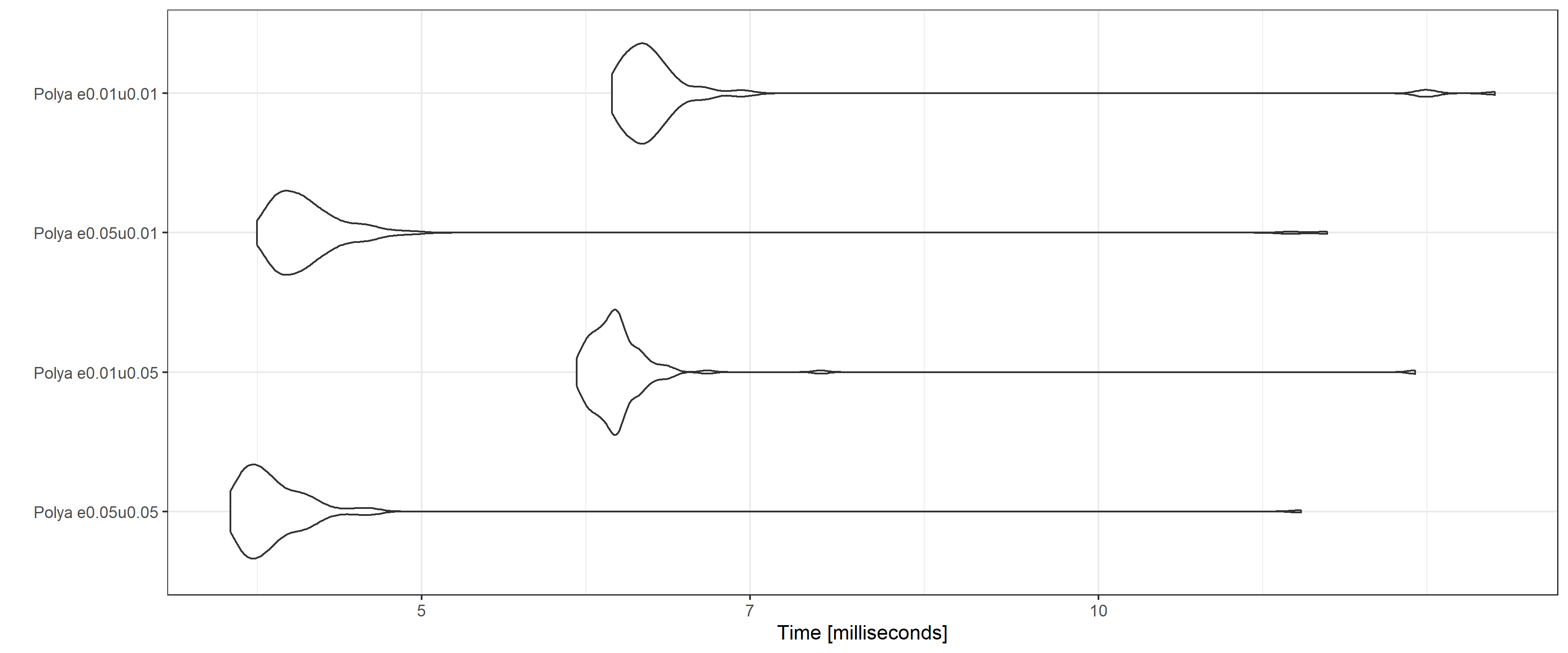}
	\caption{A violin plot of the running time for 100 runs of \lstinline{plolya(res_mdp, eps, ups)} over a grid of $\epsilon \in \{0.01, 0.05\}$ and $\upsilon \in \{0.01, 0.05\}$.}
	\label{fig:bench_pol}
\end{figure}

P\'{o}lya completion for $k = 100$ samples from such a run of \lstinline{mdp} with errors as low as $\epsilon = \upsilon = 0.01$ takes on average just over 10 milliseconds \textit{in serial}.
This puts the running time for each of the $k$ samples at around 100 microseconds, which can potentially be executed in parallel over $k$. 
For experimenters who don't have a few seconds to spare, this makes the extension of the results of Escobar \& West's Gibbs sampler nearly instantaneous. 
The running time for the \lstinline{polya} function for various values of $\epsilon$ and $\upsilon$ is shown in Figure \ref{fig:bench_pol}.

Illustrations and benchmarking for the software were conducted in a Windows environment on an Intel Core i5-8600K CPU clocked to 4.8GHz.

\section{Discussion}

Motivated by the sampling of the incomplete missing data approach for deriving full posterior summaries, including uncertainty quantification, and motivated in \cite{Fong2022}, our idea for estimating the Bayesian MDP model is to use the simplest algorithm currently available, which is that of \cite{Escobar1994}, and to then use the predictive distributions of the P\'olya--urn scheme to generate the full random distributions. While we acknowledge these could be obtained using alternative algorithms, they are   necessarily more complicated to implement, requiring the introduction of carefully chosen latent variables. That is foremost, special samplers are required and, secondly, there is no parallel aspect which is one of the key features of our method.

Putting the idea as simply as we can; we regard in the \cite{Fong2022} sense the missing data as $\theta_{n+1\infty}$ since we are easily able, via the \cite{Escobar1994} sampler, to have access to posterior samples $\theta_{1:n}$. For each such sample, we sample $\theta_{n+1:\infty}$ given $\theta_{1:n}$ via the P\'olya--urn scheme. Not only is this incredibly simple to implement, but is also incredibly fast.





\begin{thebibliography}{1}
	
	
	\bibitem{bernardo94}{\sc J.~M.~Bernardo and A.~F.~M. Smith},
	{\em Bayesian Theory},
	Wiley, (1994). 
	
	\bibitem{berti06} {\sc P.~Berti, L.~Pratelli and P.~Rigo},
	{\em Almost sure weak convergence of random probability measures},
	Stochastics (2006), 78, 91--97.
	
	\bibitem{billingsley99} {\sc P.~Billingsley},
	{\em Convergence of Probability Measures},
	J. Wiley \& Sons, Inc. New York (1999).
	
	\bibitem{Black1973} {\sc D.~Blackwell and J.~B.~MacQueen},
	{\em Ferguson distributions via P\'olya urn schemes},
	Annals of Statistics (1973), 1, 353--355. 
	
	\bibitem{Carreira2000} {\sc M.A.~Carreira-Perpinan},
	{\em Mode-finding for mixtures of Gaussian distributions},
	IEEE Transactions on Pattern Analysis and Machine Intelligence (2000), 22, 1318--1323
	
	
	\bibitem{deFinetti1937} {\sc B.~de Finetti},
	{\em Foresight: Its logical laws, its subjective sources} 
	In Annales de l’Institut Henri Poincar\'e (1937), 7, 1–-68. 
	(English translation by H.E. Kyburg, Jr. and H.E. Smokier, eds. in Studies in Subjective Probability. 
	(2nd ed. 1980), Robert E. Krieger, Huntington, New York.)
	
	\bibitem{doob49} {\sc  J.~L.~Doob},
	{\em Application of the Theory of Martingales},
	In Le Calcul des Probabilites et ses Applications 23--27. Colloques Internationaux du Centre National de la Recherche Scientifique, Paris (1949). 
	
	\bibitem{Eddelbuettel2011} {\sc  D.~Eddelbuettel},
	{\em {Rcpp}: Seamless {R} and {C++} Integration},
	Journal of Statistical Software (1949), 40, 1--18.
	
	\bibitem{Efron1979} {\sc B.~Efron},
	{\em Bootstrap methods: another look at the jackknife},
	Annals of Statistics (1979), 71, 1--26.
	
	\bibitem{Escobar1988} {\sc M.~D.~Escobar},
	{\em Estimating the means of several normal populations by nonparametric estimation of the distirbution of the means},
	Unpublished PhD dissertation (1988), Yale University.
	
	\bibitem{Escobar1994} {\sc M.~Escobar and M.~West},
	{\em Bayesian density estimation and inference using mixtures},
	Journal of the American Statistical Association (1994), 90:577--588.
	
	\bibitem{Ferguson1973} {\sc T.~S.~Ferguson},
	{\em A Bayesian analysis of some nonparametric problems},
	Annals of Statistics (1973),  1, 209--230. 
	
	\bibitem{Favaro2013} {\sc S.~Favaro, A.~Lijoi and I.~Pruenster},
	{\em Conditional formulae for Gibbs type exchangeable random partitions}, Annals of Applied Probability (2013), 23:1721--1754. 
	
	\bibitem{Ferguson1983} {\sc T.~S.~Ferguson},
	{\em Bayesian density estimation by mixture of normal distributions},
	Recent Advances in Statistics (1983), Academic Press, Inc.
	
	\bibitem{Fong2022} {\sc E.~Fong, C.~Holmes and S.~G.~Walker},
	{\em Margingale posterior distributions},
	Revised for Journal of the Royal Statistical Society, Series B (2022).
	
	\bibitem{Green1995} {\sc P.~J~.Green},
	{\em Reversible jump Markov chain Monte Carlo computation and Bayesian model determination},
	Biometrika (1995), 82:711--732.
	
	
	\bibitem{Green2001} {\sc P.~J.~Green and S.~Richardson},
	{\em Modelling heterogeneity with and without the Dirichlet process},
	Scandinavian Journal of Statistics (2001), 28: 355-375. 
	

	\bibitem{Hahn2018} {\sc P.~R.~Hahn, R.~Martin and S.~G.~Walker},
	{\em On recursive Bayesian predictive distributions},
	Journal of the American Statistical Association (2018), 113, 1085--1093.
	
	\bibitem{Hjort2010} {\sc N.~L.~Hjort, C.~Holmes, P.~Mueller and S.~G.~Walker},
	{\em Bayesian Nonparametrics},
	Cambridge University Press (2010).
	
	
	\bibitem{Ishwaran2002} {\sc H.~Ishwaran and L.~F.~James},
	{\em Approximate Dirichlet process computing in finite normal mixtures: smoothing and prior information},
    Journal of Computational and Graphical Statistics (2002), 11:508--532.
    
    \bibitem{Kalli2011} {\sc M.~Kalli, J.~E.~Griffin and S.~G.~Walker},
    {\em Slice sampling mixture models},
    Statistics and Computing (2011), 21:93--105.
    
    \bibitem{Kuo1986} {\sc L.~Kuo},
    {\em Computations of mixtures of Dirichlet processes},
    SIAM Journal on Scientific and Statistical Computing (1986), 7:60--71.
    
    \bibitem{Lo1984} {\sc A.~Y.~Lo},
    {\em On a class of Bayesian nonparametric estimates: I. Density estimates},
    The Annals of Statistics (1984), 12:351--357.
    
    \bibitem{MacEachern1994} {\sc S.~N.~MacEachern},
    {\em Estimating normal means with a conjugate style Dirichlet process prior},
    Communications in Statistics (1994), 23:727--741.
    
    \bibitem{Mena2015} {\sc R.~H.~Mena and S.~G.~Walker},
    {\em On the Bayesian mixture model and identifiability},
    Journal of Computational and Graphical Statistics (2015), 24:1155--1169.
    
   \bibitem{Tardella1998} {\sc P.~Muliere and L.~Tardella},
   {\em Aprroximating distributions of random functionals of Ferguson--Dirichlet priors},
   The Canadian Journal of Statistics (1998), 26:283--297.
    
    \bibitem{Papas2008} {\sc O.~Papaspiliopoulos and G.~O.~Roberts},
    {\em Retrospective Markov chain Monte Carlo methods for Dirichlet process hierarchical models},
    Biometrika (2008), 95:169--186.
    
    \bibitem{Pitman1997} {\sc J.~Pitman and M.~Yor},
    {\em The two parameter Poisson--Dirichlet distribution derived from a stable subordinator},
    The Annals of Probability (1997), 25:855--900.
    
    \bibitem{Richard1997} {\sc S.~Richardson and P.~J.~Green},
    {\em On Bayesian analysis of mixtures with an unknown number of components},
    Journal of the Royal Statistical Society, Series B (1997), 59:731--792.

	\bibitem{Robert2007} {\sc C.~P.~Robert},
	{\em The Bayesian Choice},
	Springer (2007).
	
	\bibitem{Rubin1981} {\sc D.~B.~Rubin},
	{\em The Bayesian bootstrap},
	Annals of Statistics (1981), 9, 130--134.
	
	\bibitem{R2022} {\sc R~Core~Team},
	{\em R: A Language and Environment for Statistical Computing},
	R Foundation for Statistical Computing (2022).
	
	\bibitem{Roeder1990} {\sc  K.~Roeder}, 
	{\em Density estimation with confidence sets exemplified by superclusters and voids in galaxies}, 
	Journal of the American Statistical Association (1990), 85, 617--624.
	
	\bibitem{Ross2022} {\sc G.~J.~Ross, D.~Markwick, K.~Mulder, and G.~Sighinolfi}
	{\em dirichletprocess: An R Package for Fitting Complex Bayesian Nonparametric Models},
	(2022)
	
	\bibitem{Sethuraman1994} {J.~Sethuraman},
	{\em A constructive definition of Dirichlet priors},
	Statistica Sinica (1994), 4:639--650.
	
	\bibitem{Stephens2000} {\sc M.~Stephens},
	{\em Dealing with label switching in mixture models},
	Journal of the Royal Statistical Society, Series B (2000), 62:795--809.
 	
 	\bibitem{Venables2002} {W.~N.~Venables B.~D.~Ripley},
 	{\em Modern Applied Statistics with S},
 	Springer (2002).
	
	\bibitem{Walker2007} {S~.G.~Walker},
	{\em Sampling the Dirichlet mixture model with slices},
	Communications in Statistics (2007), 36:45--54.
	
	\bibitem{williams91} {\sc D.~Williams},
	{\em Probability with Martingales},
	Cambridge University Press (1991).
	
	
	
\end{thebibliography}
\end{document}